\documentclass[sigconf]{acmart}
\usepackage{booktabs} 
\usepackage[linesnumbered,ruled]{algorithm2e}
\usepackage{bbm}
\usepackage{natbib}
\usepackage{multirow}
\copyrightyear{2018}
\acmYear{2018} 
\acmConference[WWW 2018]{The 2018 Web Conference}{April 23--27, 2018}{Lyon, France}
\acmBooktitle{WWW 2018: The 2018 Web Conference, April 23--27, 2018, Lyon, France}
\acmPrice{}
\acmDOI{https://doi.org/10.1145/3178876.3186021}
\acmISBN{978-1-4503-5639-8}

\begin{document}
\title{MemeSequencer: Sparse Matching for Embedding Image Macros}

\author{Abhimanyu Dubey}
\affiliation{%
  \institution{Massachusetts Institute of Technology}
}
\email{dubeya@mit.edu}

\author{Esteban Moro}
\affiliation{%
  \institution{Massachusetts Institute of Technology}
}
\affiliation{%
  \institution{Universidad Carlos III de Madrid}
}
\email{emoro@math.uc3m.es}

\author{Manuel Cebrian}
\affiliation{%
  \institution{Massachusetts Institute of Technology}
}
\email{cebrian@mit.edu}

\author{Iyad Rahwan}
\affiliation{%
  \institution{Massachusetts Institute of Technology}
}
\email{irahwan@mit.edu}
\renewcommand{\shortauthors}{Dubey et al.}

\begin{abstract}
The analysis of the creation, mutation, and propagation of social media content on the Internet is an essential problem in computational social science, affecting areas ranging from marketing to political mobilization. A first step towards understanding the evolution of images online is the analysis of rapidly modifying and propagating memetic imagery or `memes'. However, a pitfall in proceeding with such an investigation is the current incapability to produce a robust semantic space for such imagery, capable of understanding differences in Image Macros. In this study, we provide a first step in the systematic study of image evolution on the Internet, by proposing an algorithm based on sparse representations and deep learning to decouple various types of content in such images and produce a rich semantic embedding. We demonstrate the benefits of our approach on a variety of tasks pertaining to memes and Image Macros, such as image clustering, image retrieval, topic prediction and virality prediction, surpassing the existing methods on each. In addition to its utility on quantitative tasks, our method opens up the possibility of obtaining the first large-scale understanding of the evolution and propagation of memetic imagery.
\end{abstract}

%
%


\maketitle
\section{Introduction}
Social networks have increasingly become an integral part of modern life. Recent research in computational social science has focused on detecting the most shared content, the extent and pace of sharing of content, and the most influential content-sharing agents in a social network~\cite{berger2012makes, berger2011drives, barabasi2016network, buskens1999new, buskens2000cooperation}. This line of inquiry, based on the study of predicting and understanding \textit{content virality} has also risen interest in computer science~\cite{cheng2014can, valente1995network}. Content diffusion online can be understood as a product of two intertwined properties: i) the nature of the content, its evolution and mutations, and ii) the properties of the social network on which it propagates.\\ \hspace*{10pt}
Diffusion of content and \textit{cascade prediction} have received substantial attention in this domain. Several lines of recent research have focused on understanding and predicting cascades~\cite{cheng2014can}, the probabilities of information diffusion in cascades~\cite{saito2008prediction}, and the recurrence of cascades~\cite{cheng2016cascades}. These cascades are crucial in understanding the influence of the underlying social network on predicting the extent of propagation (popularity or virality) and provide strong insights into the importance of strong community structures in content propagation~\cite{weng2013virality}. Extensive research has also been done in understanding the strength and extent of online community structures and their impact on information diffusion~\cite{weng2014predicting,weng2015topicality}.\\\hspace*{10pt}
With increased online big data collection and processing, research has focused on understanding content virality through the information contained in online imagery or text~\cite{bakshy2012role, coscia2014average, gleeson2016effects}. Contrary to the earlier mentioned research, this line of focus looks at the impact of content in predicting virality, independently from the network structure and its constituent effects of social reinforcement, homophily and spreading pattern. Using computer vision techniques, studies have looked at regions of images that promote content virality~\cite{dubey2017modeling, deza2015understanding}.\\\hspace*{10pt}
An interesting combination of these two different lines of research is the study of evolution of information in social networks~\cite{adamic2016information}. Since many memes exist in the social network that persist by mutating constantly~\cite{coscia2013competition, coscia2014average}, understanding the mutations that are responsible for accelerating or hindering the popularity of a meme can be influential in content creation and understanding the cultural composition of online communities. An issue, however, with this line of study is the difficulty in isolating the \textit{underlying cultural meme} from its various manifestations in online content~\cite{knobel2007online, shifman2013memes}.\\
\begin{figure}[t]
  \centering
  \includegraphics[width=0.45\textwidth]{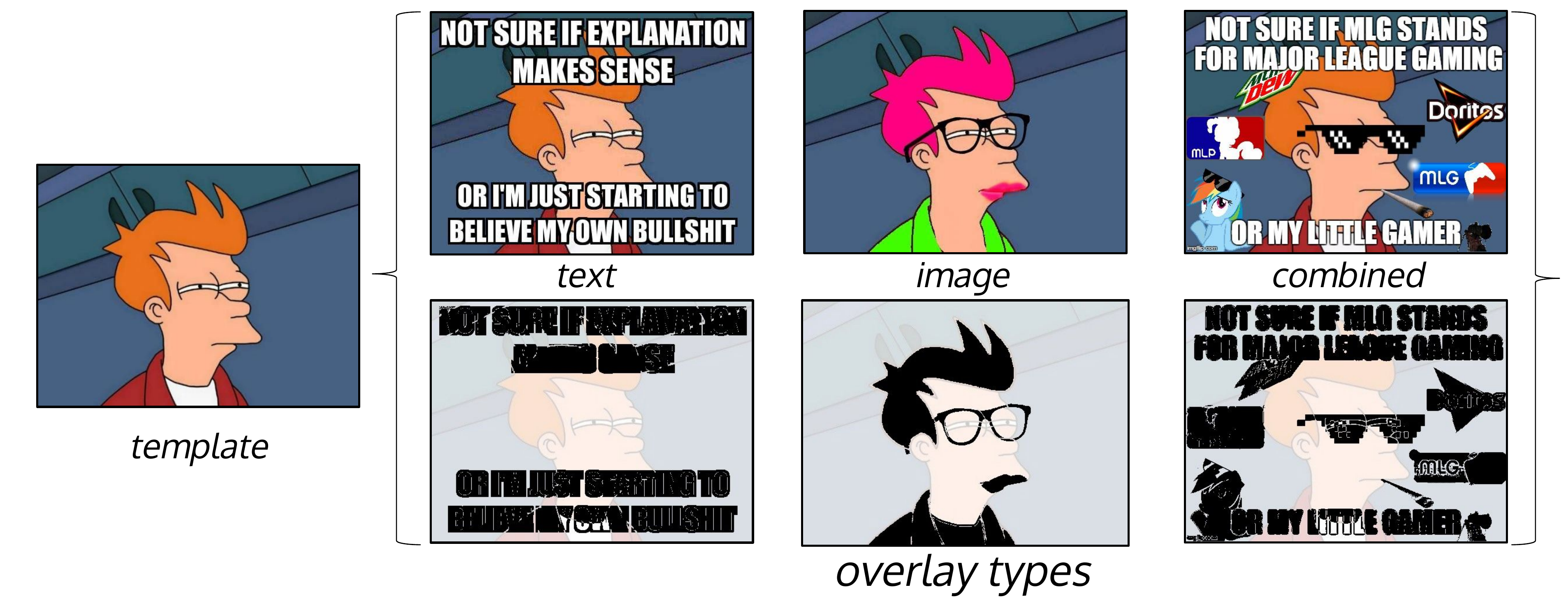}
  \caption{A sample image macro (``Futurama Fry'') with its most common overlays.}
  \label{fig:image_modifications}
\end{figure}
\hspace{10pt}
Identifying latent cultural memes from content such as tweets has been attempted first by Leskovec~\textit{et al.}\cite{leskovec2009meme}, utilizing topic modeling, without explicitly considering mutations in content. Approaches such as n-grams and deep neural representations of text~\cite{dos2014deep} have also been utilized to some success. When operating on Twitter data, hashtags provide a grounded and less noisy representation of a meme, that has been utilized in studying the propagation of associated content on social networks~\cite{romero2011differences, page2012linguistics}. The work of Coscia~\cite{coscia2013competition, coscia2014average} has studied the nature of competition between image macros online. The systematic study of mutations, however, still remains elusive under these approaches, and to the best of our knowledge, there is no work on the evolution of image-based memes.\\\hspace*{10pt}
In this work, we provide a systematic framework and associated semantic feature space to study memetic imagery. Unlike the evolution of text, images mutate and evolve in a relatively controlled manner on the Internet, typical of which is the propagation of Image Macros, the most common type of online visual meme~\cite{knobel2007online}. As described by Knobel and Lankshear~\cite{knobel2007online}, an Image Macro is the representation of an idea using an image superimposed with text optional alternative imagery. This form of representing a meme has been incredibly successful at dissemination, and is extremely popular on social networks~\cite{meme_font_impact}.\\ \hspace*{10pt}
In their most common form, Image Macros usually possess one or two lines of text flanking the template image in the center. Additionally, they may have altered imagery superimposed on the template image as well. Their etymology stems from the usage of the word ``macro'' in computer science, as a `rule or a pattern that maps an input to an output'~\cite{mcilroy1960macro}. This highlights the usage of the Image Macro as a general purpose meme representation, that can be altered to fit the context specified by the overlaid text. The combination of the overarching memetic theme provided by the instantly-recognizable template image with the subtle contextual information provided by the overlaid text or imagery creates an instantly perceivable new meme that is versatile and adapted to the targeted community, justifying its prevalence in social media.\\ \hspace*{10pt}
Scientific inquiry involving the propagation of these Image Macros can hence provide a stronger signal in understanding the transformation of cultural memes. The primary problem, which we focus on in this work, is the creation of a semantic representation for Image Macros that preserve the semantic information in each image macro while mapping different instances of image macros created from the same base template image close together, preserving the global context (supplied by the image), while additionally maintaining the individual context provided by each image macro. The baseline technique to solve a problem such as this would be to use deep neural network features, which have shown tremendous capabilities in encapsulating information from images. However, deep convolutional networks cannot decouple the overlaid imagery and template imagery, and process the overlay as noise, when the overlay in fact provides critical contextual information about the macro itself. This results in a loss of information from the macros, mapping most of them to the similar representation, which only amplifies as more overlays are made.\\ \hspace*{10pt}
In this study, we create an algorithm that first uses the idea of sparse representation to identify template images from each Image Macro, and then using the obtained template, decouples the overlaid information from the base template. We then proceed to extract multimodal features from each image, resulting in a rich, informative and robust feature representation. Using this feature representation, we demonstrate remarkable improvements across several qualitiative and quantitative tasks involving social media imagery, demonstrating the conceptual and functional superiority of our approach from other baseline techniques. In cases where the template set is not known beforehand, we also provide an algorithm that can recover the template image from a set of sample macros based on median blending of images.
\section{Method}
Our method is based on a strong underlying assumption -- memetic imagery online contains substantial amounts of Image Macros, that are constructed from a set of template images by choosing a template image and overlaying text and/or additional imagery on it. This assumption is exploited in our formulation. We begin the algorithmic description with preliminaries:\\ \hspace*{10pt}
\textbf{Target Set}: Our target set $\mathbf T$ is the set of images that we wish to embed in a semantically-grounded space. In our experiments, this usually is the dataset that we conduct experiments on, scraped from websites such as Memegenerator~\cite{coscia2014average} or Quickmeme~\cite{coscia2013competition}. We use this set of images to construct the set of template images, following Algorithm~\ref{algo:template_set_construction}, that we then use for Sparse Matching and feature extraction.\\ \hspace*{10pt}
\textbf{Template Set}: The template set $\mathbf S$ is the set of images with no overlays that we match each image in the target set with, to obtain the decoupled \textit{Image Macro} representation. This template set can be supplied beforehand, but in case it is not, we construct it from the Target Set itself using an algorithm involving Sparse Matching and Median Blending.\\ \hspace*{10pt}
\textbf{Overlay Types}: Figure~\ref{fig:image_modifications} specifies the typical kinds of overlays on template images to produce Image Macros. The most common overlay is simple text in a white font (as shown in the section `text overlay'). There can be modifications in the color or the addition of an image, which fall under the category of `image overlay'. Additionally, both these things may be present together, images of which fall in the `combined overlay' category. We take care of minute variations in color across images by contrast normalization.\\ \hspace*{10pt}
The goal of our Sparse Matching algorithm is to obtain the template image any sample \textit{Image Macro} has been constructed from. Using this template image, we can then decouple the image overlay from the Macro, and process the overlay and corresponding template separately to decouple the local context (specified by the overlay) and the global context (specified by the template).
\subsection{Algorithm Overview}
Our task can be summarized as learning an embedding function $f : \mathbb R^n \rightarrow \mathbb R^d$ that maps images to a low-dimensional embedding that preserves semantic content. To create an embedding, our method follows three distinct subroutines as described below:\\
\hspace*{10pt} (1) \textbf{Overlay Decoupling}: The first step of the algorithm is to identify and separate the overlaid content from the template image. To do this we employ global image contrast normalization followed by $\ell^1$-sparse reconstruction, first introduced in the seminal work on face recognition by Wright et al.~\cite{wright2009robust}. \\
\hspace*{10pt} (2) \textbf{Image Feature Extraction}: Image features learnt through deep convolutional neural networks (CNNs) for object classification have been shown to be tremendously powerful at capturing semantic information, and have excelled at a variety of inference tasks~\cite{donahue2014decaf}. To capture the semantic content of the imagery, we use deep CNNs trained on image classification, and then finetuned on our target task. \\
\hspace*{10pt} (3) \textbf{Textual Feature Extraction}: To augment the information provided by the image features, we additionally extract text present in the images using optical character recognition (OCR), and augment our embedding with features extracted from this text data. To learn these text features, we use a deep recurrent neural network~\cite{medsker2001recurrent}, inspired by their immense success in a variety of inference tasks in the domain of natural language processing~\cite{collobert2008unified, graves2005framewise, zheng2013deep}.\\ \hspace*{10pt}
After overlay decoupling and multimodal feature extraction, we concatenate the obtained text and image features to produce a powerful embedding for image retrieval and clustering of the target meme imagery, whose representational strength we verify in several experiments as described later.
\subsection{Decoupling Overlays from Templates}
The first subroutine of our method involves decoupling the image overlay from the template it was produced from. This can be done using a straightforward pixel-wise subtraction, however, we do not have the source template of each image \textit{a priori}, which makes the first task the identification or matching of the correct template image from the provided test image. We begin by first normalizing the color in the image with a global pixel-wise mean normalization, to remove the slight aberrations in color across images. We then downsample each image in both the template image set $\mathbf S$ and target image set $\mathbf T$ to a fixed resolution of $48 \times 48$ pixels. Consider these downsampled sets as $\mathbf S_d$ (template set) and $\mathbf T_d$ (target set) respectively. Given these sets of normalized, downsampled images, we now describe the sparse representation algorithm.
\subsubsection{Sparse Representation}
Introduced by Wright et al.~\cite{wright2009robust}, the sparse representation algorithm provides a framework for inference under the assumption that the training samples lie on a subspace. Hence, each input test point can be written as a sparse linear combination of the training points. If we consider the training set to be the set of downsampled template meme images $\mathbf S_d$, and test set to be each image in the target set $\mathbf T_d$, we can apply the sparse representation algorithm to match each sample in $\mathbf T_d$ to a sample in $\mathbf S_d$. By this process, we can effectively recover the original template the macro was created from, and decouple the template from the overlay.\\ \hspace*{10pt}
\textbf{Matching}: Let the total number of images present across all templates be $m$, and images in each class $i$ be given by $m_i$. Hence, $\sum_{i=1}^k m_i  = m$. Given a set $\mathbf S_{d,i}$ of $m_i$ images, represented as a matrix $[\mathbf s_{1,i}, ..., \mathbf s_{m_i,i}] \in \mathbb R^{n \times m_i}$ belonging to class $i$, any new target sample $\mathbf y \in \mathbf T_d \subset \mathbb R^n$ belonging to template $i$ will approximately lie in the linear span of the training samples of $\mathbf S_{d,i}$. Hence:
\begin{align}
\mathbf y &= \alpha_{1,i} \mathbf s_{1,i} + \alpha_{2,i} \mathbf s_{2,i} + \alpha_{3,i} \mathbf s_{3,i} + ... + \alpha_{m_i,i} \mathbf s_{m_i,i} \\ \intertext{Where $\alpha_{i,j} \in \mathbb R$. If we consider the set of all classes $\mathbf S_d = \cup_i^k \ (\mathbf S_{d,i})$, we can write the matrix for all the $m_i$ samples from each of the $k$ classes as:}
\mathbf A &:= [\mathbf S_{d,1}, \mathbf S_{d,2}, ..., \mathbf S_{d,k}] = [\mathbf s_{1,1}, \mathbf s_{2,1},...,\mathbf s_{m_k, k}] \\ \intertext{Using this, we can write the linear representation of $\mathbf y$ over all training samples as:}
\mathbf y &= \mathbf A \mathbf x_0 \ \in \mathbb R^n
\end{align}
where, $\mathbf x_0 = [0,...,0,\alpha_{1,i},\alpha_{2,i},...,\alpha_{m_i,i}, 0, ..., 0]^\top \in \mathbf R^m$ is a coefficient vector with zero entries except for samples belonging to class $i$. As proposed in \cite{wright2009robust}, to obtain the sparsest solution of the above equation, we have to solve the following $\ell^0$ optimization problem:
\begin{align}
\hat{\mathbf x}_0 = \arg \min \lVert \mathbf x \rVert_0 \ \ \ \text{, subject to} \ \ \ \mathbf A \mathbf x = \mathbf y
\end{align}
This corresponds to finding a coefficient vector $\hat{\mathbf x}_0$ that contains the smallest number of non-zero entries. This problem in its current form is to find the sparsest solution of an underdetermined linear system, and is NP-hard, and even difficult to approximate~\cite{amaldi1998approximability}. As proposed in~\cite{donoho2006most, candes2006stable, wright2009robust}, under certain assumptions of sparsity, the $\ell^0$ minimization solution is equivalent to the $\ell^1$ minimization solution. We therefore solve the following problem:
\begin{align}
\hat{\mathbf x}_1 = \arg \min \lVert \mathbf x \rVert_1 \ \ \ \text{, subject to} \ \ \ \mathbf A \mathbf x = \mathbf y
\end{align}
This corresponds to finding a coefficient vector $\hat{\mathbf x}_0$ that has the minimum $\ell^1$-norm ($\sum_{i=1}^m |\mathbf x^{(i)}_i|)$. Using standard linear programming methods, one can solve this problem in polynomial time~\cite{chen2001atomic}. As described in detail in~\cite{wright2009robust}, this algorithm recovers the correct training class even in the presence of severe occlusion (as provided by the overlaid text and/or pictures to an image), and heavy amounts of random noise. Hence, this method is ideal in our application case.\\
\begin{algorithm}
\caption{Template Matching via Sparse Representation}
\label{algo:image_correction}
\SetKwInOut{Input}{Input}
\SetKwInOut{Output}{Output}
\Input{Template Set $\mathbf S_d$, Target Set $\mathbf T_d$, threshold $t_r$, no. of different templates $k$}
\Output{Matched Set $\mathbf O_d$}
Set $\mathbf O_d \leftarrow \emptyset$\\
Compute $\mathbf A \leftarrow [\mathbf s_{1,1}, \mathbf s_{2,1},...,\mathbf s_{m_k, k}]$ \label{algo_part:compute_a}\\
\For{Image $\mathbf t_i$ in $\mathbf T_d$}{
    $\hat{\mathbf x} \leftarrow \arg \min \lVert \mathbf x \rVert_1 \ \ \ \text{, subject to} \ \ \ \mathbf A \mathbf x = \mathbf t_i$\\
		\uIf{$\lVert \mathbf A \hat{\mathbf x} - \mathbf t_i \rVert_2 \leq t_r$}{
	    \For{Class $c$ from $1$ to $k$}{
	    	Compute $z_c \leftarrow \sum_{j=1}^{m_c} \mathbbm 1 \{\hat{\mathbf x}_{j,c} > 0\}$
	    }
	    Set $\mathbf z^{(i)} \leftarrow [z_1, z_2, z_3, ..., z_k]$ \\
	    Compute $\hat{z}^{(i)} \leftarrow \arg \max ( \mathbf z^{(i)})$\\
	    Set $\mathbf O_d \leftarrow \mathbf O_d \cup \{\mathbf s_{1,\hat{z}^{(i)}}\}$
			}
		\Else{
			Set $\mathbf O_d \leftarrow \mathbf O_d \cup \{\mathbf t_i\}$
		}
    }
\Return{$\mathbf O_d$}
\end{algorithm}
\begin{figure}[t]
  \centering
  \includegraphics[width=0.45\textwidth]{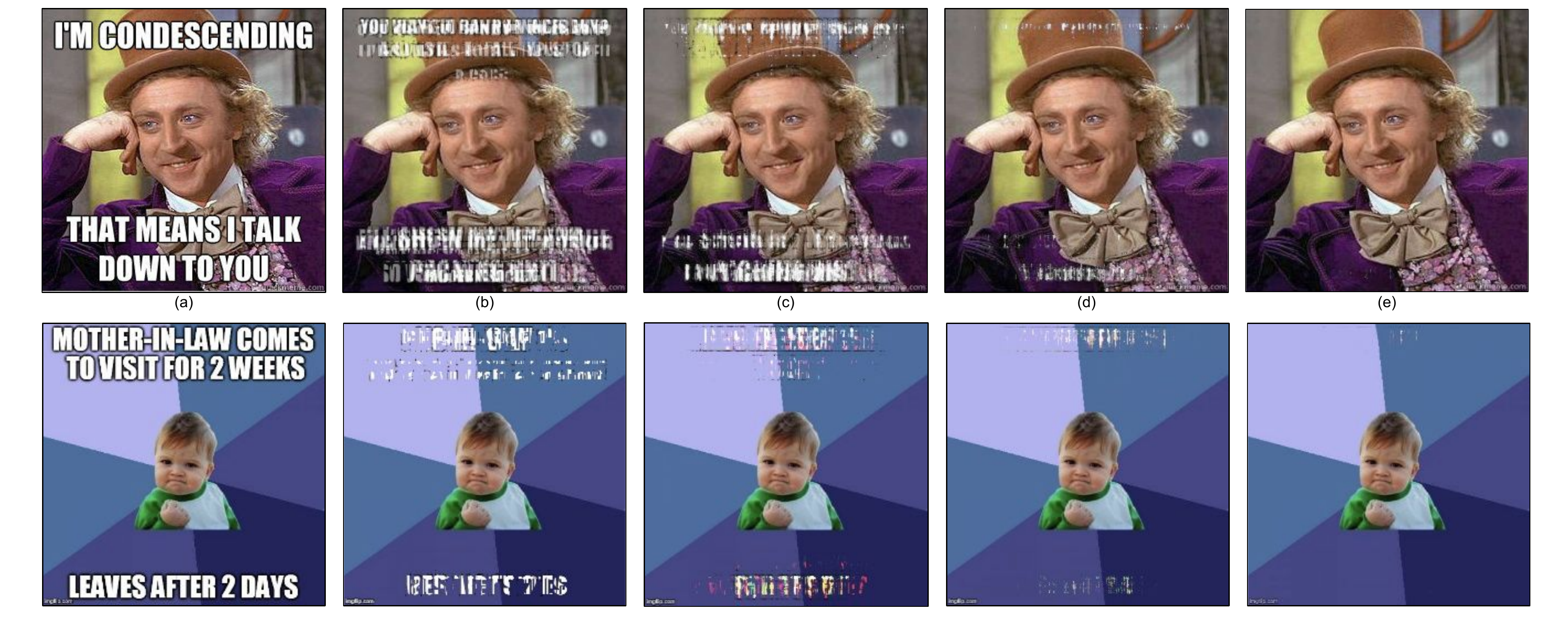}
  \caption{\small Sample recovery of the template images ``Willy Wonka'' and ``Success Kid'' from the target set examples. We show the obtained template at (a) 0 iterations, (b) 10 iterations, (c) 20 iterations, (d) 50 iterations and (e) 100 iterations.}
	\label{fig:median_blending}
\end{figure}
Our algorithm is described in Algorithm \ref{algo:image_correction}. We begin by computing our downsampled template and target sets $\mathbf S_d$ and $\mathbf T_d$ respectively, and set the output set $|mathbf O_d$ to the empty set. For each image $\mathbf t_i$ in the target set, we first compute the sparse representation weights $\mathbf x$ as described earlier. Using the components of the sparse representation, we then proceed to evaluate the number of non-zero weights present for samples of each template class and store it in variable $z_c$ for template class $c$. Owing to the sparse representation formulation, we find that only the matched template class possesses non-zero components. We then assign the template class by choosing the class with the maximum number of non-zero sparse weights, and assign the blank template image as the corrected set for the input sample $\mathbf t_i$. In case no matching template is found (the error in reconstruction is larger than a threshold $t_r$, that is $\lVert \mathbf A \hat{\mathbf x} - \mathbf t_i \rVert_2 > t_r$), we return the target image itself as the matched template.\\ \hspace*{10pt}
Once we have the assigned template image $\mathbf s_{1,\hat{z}^{(i)}}$, we can decouple the overlay from the template by computing the decoupled overlay as $\mathbf t_i - \mathbf s_{1,\hat{z}^{(i)}}$, which can then help us do decoupled feature extraction. A caveat of this algorithm is that it requires an exhaustive template set to filter images successfully. However, in practical cases, we might not have the template set $\mathbf S$, in which case the next algorithm to construct the template image set $\mathbf S$ from the target image set  $\mathbf T$ can be used.\\ \hspace*{10pt}
\begin{algorithm}
\caption{Template Set Construction}
\label{algo:template_set_construction}
\SetKwInOut{Input}{Input}
\SetKwInOut{Output}{Output}
\Input{Target Set $\mathbf T_d$, thresholds $t_r, t_b$}
\Output{Template Set $\mathbf S_d$}
Set $\mathbf S_{d,t} \leftarrow \emptyset, \mathbf S_d \leftarrow \emptyset$\\
\For{Image $\mathbf t_i$ in $\mathbf T_d$}{
Set $\mathbf U_{\mathbf t_i} \leftarrow \emptyset$ \\
Set $\mathbf c_{\mathbf t_i} \leftarrow 0$ \\
	\uIf{$\mathbf S_{d,t} = \emptyset$}{
		Set $\mathbf S_{d,t} \leftarrow \mathbf S_{d,t} \cup \{\mathbf t_i\}$ \\
		Set $\mathbf U_{\mathbf t_i} \leftarrow \{\mathbf t_i \}$
	}
	\Else{
		Compute $\mathbf A$ from $\mathbf S_d$ (from Algorithm \ref{algo:image_correction} step \ref{algo_part:compute_a}) \\
		$\hat{\mathbf x} \leftarrow \arg \min \lVert \mathbf x \rVert_1 \ \ \ \text{, subject to} \ \ \ \mathbf A \mathbf x = \mathbf t_i$\\
		\uIf{$\lVert \mathbf A \hat{\mathbf x} - \mathbf t_i \rVert_2 \leq t_r$}{
			Compute $\mathbf s_{\hat{z}^{(i)}} \in \mathbf S_d$ (from Algorithm 1) \\
			Set $\mathbf U_{\mathbf s_{\hat{z}^{(i)}}} \leftarrow \mathbf U_{\mathbf s_{\hat{z}^{(i)}}} \cup \{\mathbf t_i \}$ \\
			\uIf{$c_{s_{\hat{z}^{(i)}}} =0$}{
				Set $\mathbf v \leftarrow$ PixelWiseMedianBlending($\mathbf U_{\mathbf s_{\hat{z}^{(i)}}}$)\\
				\uIf{$\lVert \mathbf v - \mathbf s_{\hat{z}^{(i)}} \rVert_2 \leq t_b$}{
					Set $c_{s_{\hat{z}^{(i)}}} = 1$
					}
				Set $\mathbf s_{\hat{z}^{(i)}} \leftarrow \mathbf v$
				}
		}
		\Else{
			Set $\mathbf S_{d,t} \leftarrow \mathbf S_{d,t} \cup \{\mathbf t_i\}$ \\
			Set $\mathbf U_{\mathbf t_i} \leftarrow \{\mathbf t_i \}$
		}
	}
}
\For {Image $\mathbf s_i$ in $\mathbf S_{d,t}$}{
	Set $\mathbf S_d \leftarrow \mathbf S_d \cup $ Augment($\mathbf s_i$)
}
\Return{$\mathbf S_d$}
\end{algorithm}
\textbf{Creation of Template Set}: To construct a template set automatically from the target set, we describe an algorithm that utilizes the concept of median blending in image processing~\cite{zhang2014personal}. Median Blending is a commonly used technique in image processing to obtain a stable image from several noisy images. The central idea is to iteratively grow the template set and refine the template via successive median blending.\\ \hspace*{10pt}
\begin{figure*}[t]
  \centering
  \includegraphics[width=\textwidth]{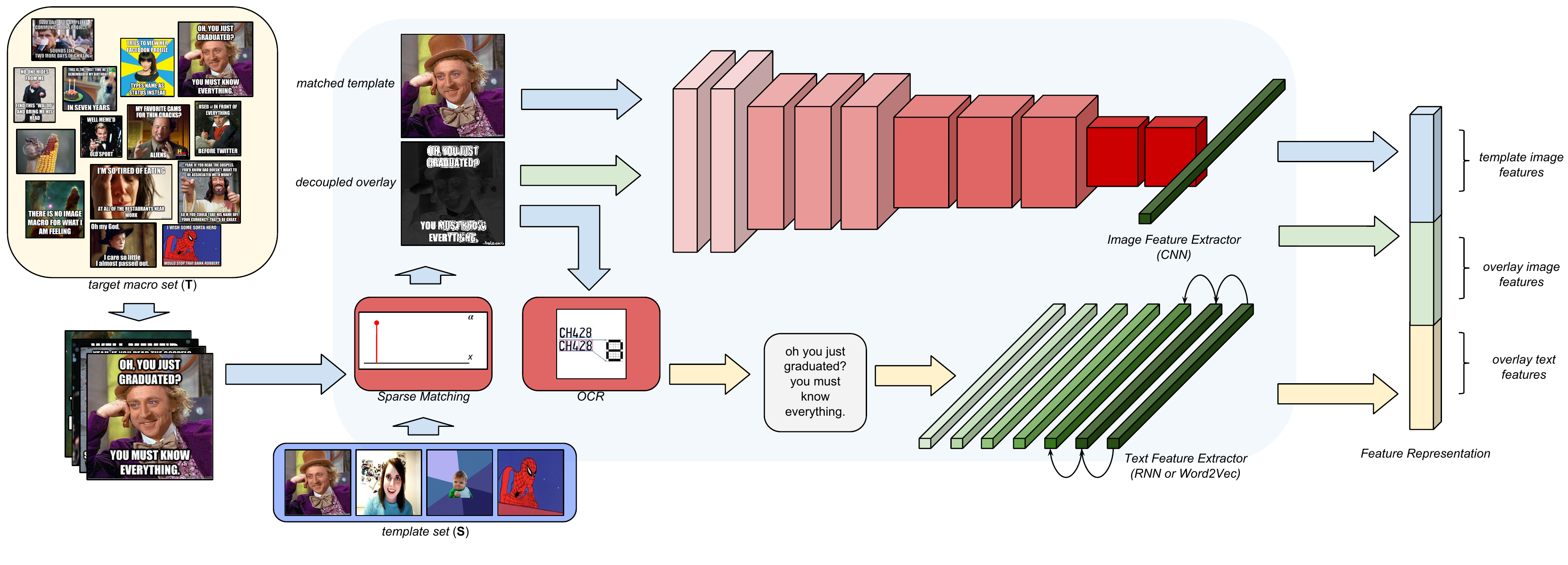}
  \caption{The central feature extraction pipeline.}
	\label{fig:main_pipeline}
\end{figure*}
We begin with an empty pre-augmented template set $\mathbf S_{d,t}$, and iterate over the target set. For the first iteration, we simply add the input image itself to $\mathbf S_{d,t}$, since our template set is empty. From here on, we maintain a set of template images $\mathbf U_{\mathbf t_i}$ for every template $\mathbf t_i$ we identify (hence, for the first iteration, we add the first image to the set $\mathbf U_{\mathbf t_i}$). For every subsequent image, we compute the sparse coefficients $\mathbf s_{\hat{z}^{(i)}}$ using $\mathbf S_{d,t}$, and if the input image matches with a template (even if a template is garbled, the sparse representation will ensure a sparse match, which we evaluate via the reconstruction error $\lVert \mathbf A \hat{\mathbf x} - \mathbf t_i \rVert_2$). If a match is found, we add the input image to the set of images corresponding to the matched template image ($\mathbf U_{\mathbf s_{\hat{z}^{(i)}}}$). We construct the new version of the template by blending all of the images in the set of matched images corresponding to that template. This blending is done by creating a new image where each pixel is the median of the corresponding pixels in all the images (referred to as PixelWiseMedianBlending).  For every input image, we proceed in a similar manner until the new obtained median image is within a small error of the median image in the previous iteration (we check if $\lVert \mathbf v - \mathbf s_{\hat{z}^{(i)}} \rVert_2 \leq t_b$, if yes, we reach convergence for image $\mathbf s_{\hat{z}^{(i)}}$, and set $c_{\mathbf s_{\hat{z}^{(i)}}} = 1$). After convergence, we do not alter the template image. Figure~\ref{fig:median_blending} describes how the median image evolves with increasing iterations.\\ \hspace*{10pt}
Once we have passed through all images in the target set, we augment the produced template set by random flips and crops of each template image (procedure described as `Augment'). This is done since several Image Macros are also created from flipped and cropped versions of the template image, and this method ensures that every test image is mapped to a template correctly. This algorithm is described in Algorithm~\ref{algo:template_set_construction}.
\subsection{Image Feature Extraction}
Once we have completed the first procedure involving decoupling the overlay, we are ready to extract features that encapsulate the semantic visual content present in the imagery. A wide variety of image features have been experimented with in computer vision literature pertaining to web and social media~\cite{deza2015understanding, dubey2017modeling, khosla2014makes}. An emerging consensus in computer vision has been the extreme efficiency of deep neural network features in capturing semantic content for classification and understanding, as described in Donahue et al.~\cite{donahue2014decaf}. From this, we continue all image feature extraction using on deep neural network models.
\subsubsection{Convolutional Neural Networks} The immense popularity of convolutional neural networks across a wide variety of computer vision tasks have made it our default choice for feature extraction. Following standard practice in computer vision~\cite{deza2015understanding, dubey2017modeling}, we consider neural network models trained on the image classification dataset ImageNet~\cite{deng2009imagenet}, and additionally consider models first trained on ImageNet followed by further fine-tuning on virality prediction datasets.
\subsubsection{Decoupled Feature Extraction} For each image $\mathbf t_i$ in the target set $\mathbf T$, we extract features from two separate images. We first extract features from the matched image from the template set $\mathbf o_i \in \mathbf O$, and then compute the difference image $\mathbf d_i = \mathbf t_i - \mathbf o_i$, and extract features from this image as well, eventually concatenating the two sets of features to form our final feature vector $\mathbf v_i$. Hence,
\begin{align}
	\mathbf v_i = [f(\mathbf o_i), f(\mathbf t_i - \mathbf o_i)]
\end{align}
Here, $f()$ is a function that maps an image to a multidimensional semantic feature vector, and is obtained from a CNN. Since the basic overlaying of text or additional imagery would be treated as noise by the CNN, we separate the two components and extract features separately. This ensures that images belonging to the same base template have features that are closer together in semantic space.
\subsection{Textual Feature Extraction} Since most memetic imagery possess modfications in the form of overlaid text, we can exploit them to produce stronger representations by a text extraction pipeline. For this, we first run Optical Character Recognition(OCR) to obtain the text (if any) contained in the image. Following this, we extract deep neural network features based on standard practices in the field of natural language processing using deep learning~\cite{yang2016hierarchical, zhang2015sensitivity, mikolov2013distributed}.\\
\textbf{Optical Character Recognition:} We use the Tesseract~\cite{smith2007overview} package for OCR, and trim excess trailing text.
\subsubsection{Word2Vec Pooling} Here we extract word2vec~\cite{mikolov2013distributed} representations (following  ~\cite{zhang2015sensitivity, mostafazadeh2016corpus, yang2016hierarchical}) for each word present in the sentence, and to produce the final representation, we average over the individual word2vec representations. The word2vec implementation used is GenSim~\cite{rehurek2011gensim} with dimensionality 1000.
\subsubsection{Skip-Thought Vectors} Kiros et al.~\cite{kiros2015skip} introduce Skip-Thought Vectors, a generic encoding scheme for sentences. The immense versatility of skip-thought vectors on a wide variety of sentence classification tasks makes them a good choice for sentence embedding creation. To extract the skip-thought (ST) features, we simply supply the extracted text to the skip-thought model, and extract the penultimate layer features.\\ \hspace*{10pt}
If in the OCR phase, we do not find any text present in the image, we later replace the features with the mean text feature for that template across all images in the Target Set, in order to minimize the impact on nearest neighbor retrieval and classification. We provide ablation and comparative studies to assess the individual performance of each of the 2 abovementioned techniques in a variety of classification tasks, and find impressive results across the board, as summarized in the experiments section.\\
\begin{figure*}[t]
  \centering
  \includegraphics[width=0.45\textwidth]{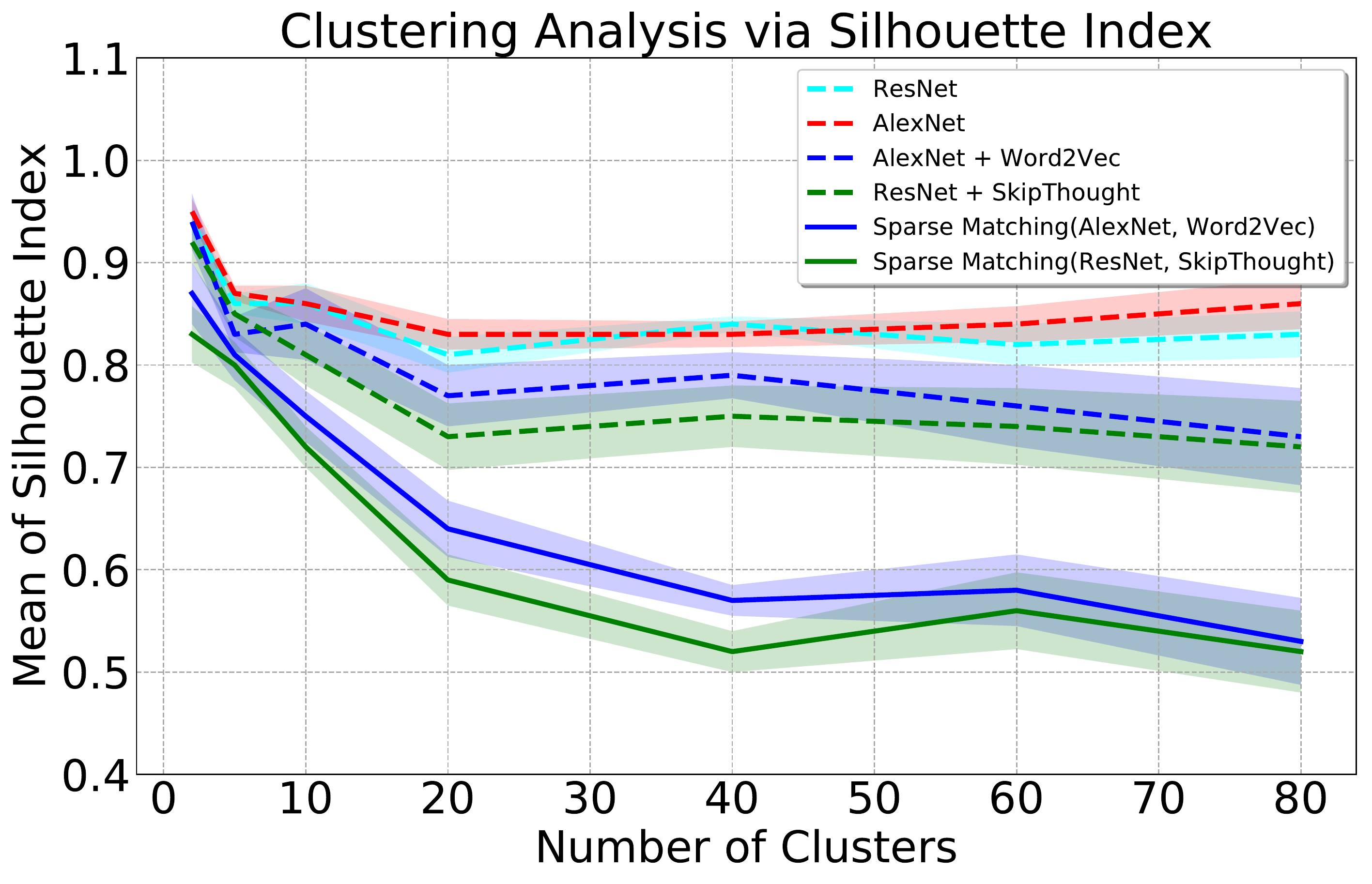}\hspace{10pt}
  \includegraphics[width=0.45\textwidth]{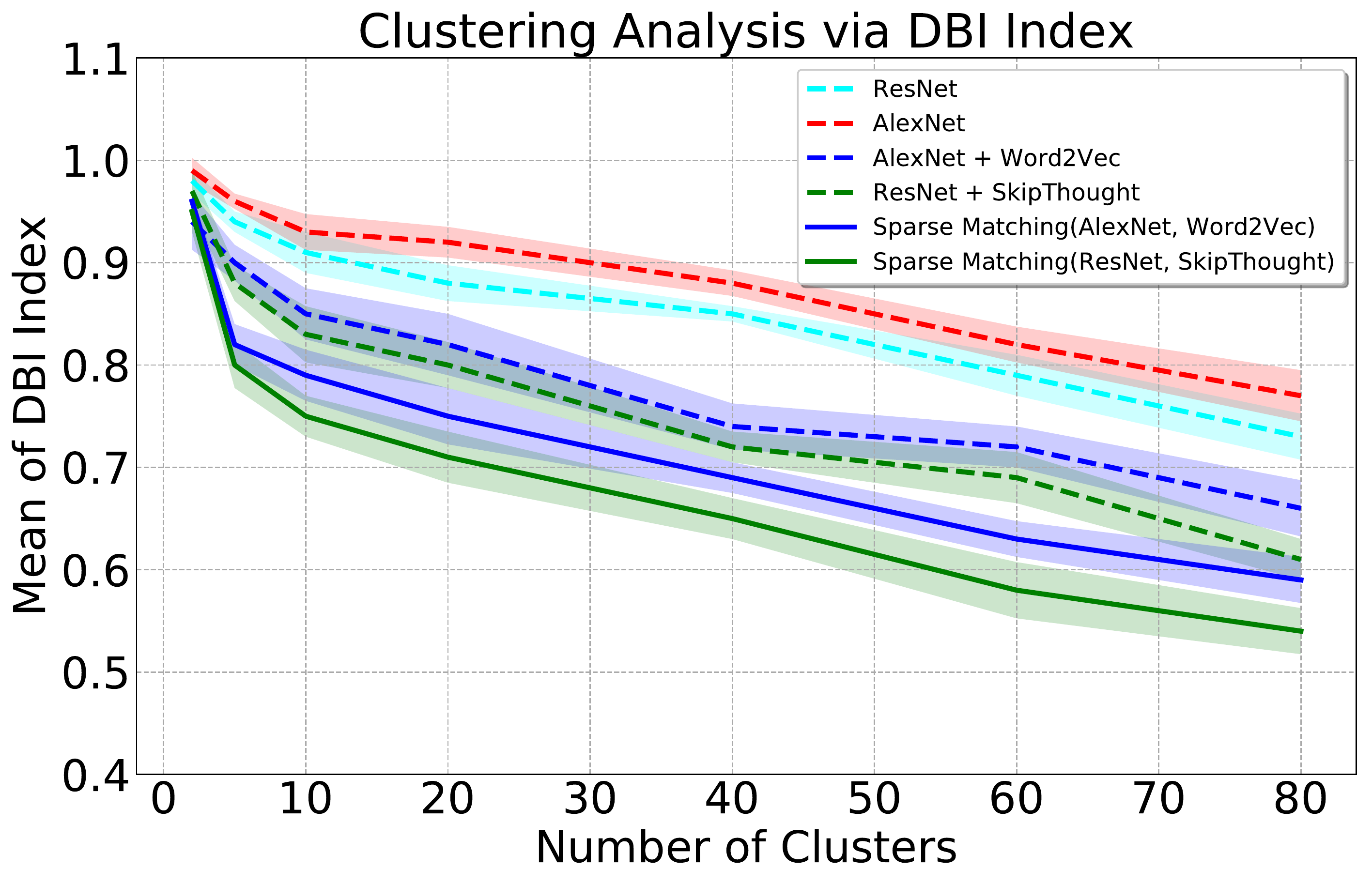}
  \caption{Evaluation of Sparse Matching on Clustering - variation of Silhouette Score (left), and Davies-Bouldin Index (right).}
	\label{fig:clustering_analysis}
\end{figure*}
\textbf{Algorithm Summary}: After image correction and individual extraction of text and image features, as described in Figure ~\ref{fig:main_pipeline}, we obtain an informative and flexible feature representation, that preserves semantic content in the presence of overlaid text and/or imagery. We analyse the effectiveness of our obtained representation across a variety of experiments, as described in the next section.
\section{Evaluation}
\subsection{Experimental Setup}
To showcase the effectiveness of our representation for memetic imagery, we perform a variety of qualitative and quantitative assessments. For all experiments, we use a setup involving NVIDIA TITAN X GPUs, and our implementation is done using the PyTorch~\cite{paszke2017pytorch} and TensorFlow~\cite{tensorflow} frameworks. For each of the individual language and vision models, we use their publicly available weights and implementations (we use GenSim~\cite{rehurek2011gensim} for Word2Vec).\\ \hspace*{10pt}
The Sparse Matching technique has two components - the image feature extractor (denoted as $I$), and the text feature extractor (denoted as $T$), giving us results of the form \textbf{Sparse Matching}$(I,T)$ for that particular choice of feature extractors. For each experiment we conduct, we vary both the feature extractors to create different variants of the algorithm, which are all compared with (i) baseline feature extractors, (ii) previous state-of-the-art algorithms on our target tasks, (iii) each feature extractor individually and (iv) the naive combination of both feature extractors, providing an exhaustive set of comparative techniques. The CNN feature extractors we use are AlexNet~\cite{krizhevsky2012imagenet}, VGGNet-16~\cite{simonyan2014very}, and ResNet-18~\cite{he2016deep}. As our results depict, the more powerful image feature extractors(ResNets and VGGNet-16) provide consistently higher performance.\\ \hspace*{10pt}
The target datasets for our task are datasets pertaining to memetic imagery, typically scraped from websites such as Memegenerator, Reddit or Quickmeme. There has been interest from both the computer vision and computational social science communities in understanding such imagery, and we use the popular datasets used in these studies, with a total of 5 test splits used:\\
(1) \textbf{Viral Images Dataset~\cite{deza2015understanding}}: Introduced in~\cite{deza2015understanding}, the Viral Images Dataset is a set of viral images filtered from the original dataset collected by Lakkaraju et al.~\cite{lakkaraju2013s} from Reddit over a period of 4 years. The dataset contains 10,078 images along with a metric for virality from 20 different image categories. This dataset contains 3 testing splits of data - (i)\textbf{Viral-Complete}(VCom), which is a collection of randomly selected pairs of images, (ii)\textbf{Viral-Pairs}(VPair), which is a collection image pairs where an image from the top 250 most-viral images is paired with an image from the top 250 least viral images, and (iii) \textbf{Viral-RandomPairs}(VRPairs), which is a set of image pairs, where one image is sampled from the top 250 most viral images, and the other is sampled at random. We utilise the predetermined training, validation and test splits.\\
(2) \textbf{Memegenerator Dataset ~\cite{coscia2014average}}: This dataset is a collection of meme images scraped from the website Memegenerator, from a period of June 27th, 2013 to July 6, 2013, accompanied with the number of upvotes for each meme as well. With a total of 326,181 images, this is the largest meme dataset, of which we use a random 70\% for training, 10\% images for validation and 20\% for testing.\\
(3) \textbf{Quickmeme Dataset ~\cite{coscia2013competition}}: Similar to the Memegenerator Dataset, this dataset is scraped from the website Quickmeme from October 2012. This dataset has a total of 178,801 images, and here as well, we use an 70-20-10 train-test-val split.\\
The end product of our algorithm is a versatile and efficient feature representation which can be potentially useful for several inference tasks involving online imagery, especially those that are propagated through social media, such as memes. To evalaute the efficacy of our proposed algorithm, the most natural initial experiment would be the image clustering and retrieval. A common task online would be phrased as ``given a certain test image $\textbf I$ and image set $\mathcal S_{\mathbf I}$, what are the images in $\mathcal S_{\mathbf I}$ that are most similar (in semantic content) to the test image $\textbf I$?'', and indeed, this is the first target task we design our algorithm for, and to evaluate our algorithm on this task, we look at three separate qualitative and quantitative experiments, as summarized below:
\subsection{Image Clustering and Retrieval}
\begin{figure}[t]
  \centering
  \includegraphics[width=0.45\textwidth]{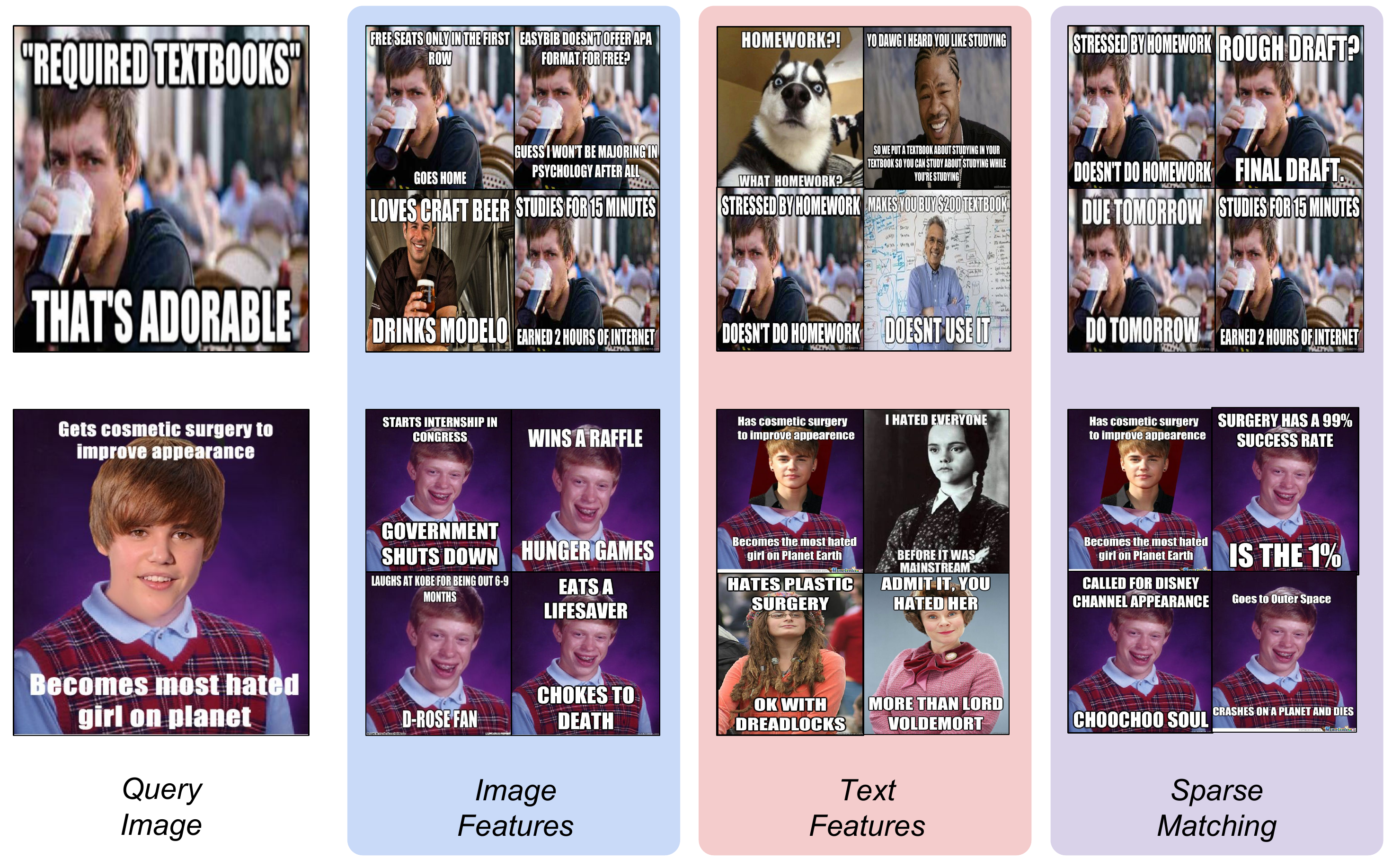}
  \caption{4-Nearest neighbour image retrieval results for two sample queries.}
	\label{fig:image_retrieval}
\end{figure}

\textbf{Clustering Analysis}: A summarizable version of the retrieval task is the task of image clustering  - understanding how our feature representation embeds the space of images. Considering that these images form smaller subsets in semantic space, we would expect distinct clusters being formed in the embedding space. Since we do not necessarily have labels for classes in most of our datasets, to assess the quality of clustering we first compare the \textit{internal clustering metrics} for our representation with the existing state-of-the-art and baseline methods. The Silhouette Score~\cite{petrovic2006comparison}(SS) and the Davies-Bouldin Index~\cite{davies1979cluster}(DBI) are two popular internal metrics used to quantifiably determine the quality of clustering of a representation. We proceed by clustering points from the Memegenerator Dataset~\cite{coscia2013competition} and vary the number of desired clusters, to obtain the clustering indices for each feature extraction algorithm as functions of the number of clusters required. Here, we use the clustering algorithm K-means, with an off-the-shelf implementation in Scipy~\cite{scipy}.\\ \hspace*{10pt}
The desiderata for efficient clustering are separable clusters, with no significant overlap, and high similarity within clusters. Or, put succintly, high intra-cluster similarity, and low inter-cluster similarity. If we consider the Silhouette Score (SS), for clusters that are balanced in size and fairly separable, we desire a low mean value of SI across clusters, and a low deviation of SS across clusters as well - representing the average deviation of cluster items from the cluster mean. Figure~\ref{fig:clustering_analysis}a summarizes the results for SS analysis, which demonstrates the distinctive clustering our algorithm provides. Similarly, for the metric DBI (Figure~\ref{fig:clustering_analysis}b), we consider the Euclidean distance as the distance metric, and observe consistently lower values, consolidating the efficiency of our algorithm in creating rich feature representations.\\ \hspace*{10pt}
\textbf{Image Retrieval and Visualization}: The previous set of results provide a quantiative summary of the improved clustering provided by our algorithm, but do not make precise an intuition to validate our representation. To strengthen the qualitative intuition behind the performance of our algorithm, we visualize using a nearest neighbor retrieval.\\ \hspace*{10pt}
We provide a set of example image retrieval queries, and compare the retrieved images for the nearest neighbours of the query in feature space. As summarized in Figure~\ref{fig:image_retrieval}, the sparse matching algorithm is robust to changes in text information, and by decoupling the text, base template and overlaid imagery, we see a richer embedding that preserves these changes in semantic space, whereas other algorithms simply treat an aberration or overlaid text as noise, providing poorer retrieval results. Just simply using image features retrieves visually similar images, yet, their text content is dissimilar. Using simply text features proceeds with the reverse effect, with various imagery being returned. Our method, however, returns a semantically aligned set of images.\\ \hspace*{10pt}
Observing the robustness of our representation and the qualitiative improvements in image retrieval, a natural extension would be understanding the performance of the feature representation on more complex tasks involving online imagery, such as inferring the popularity, content and timeline of an image. These quantitative evaluation tasks are motivated by several observations on online imagery and online content itself. It has long been established in social science literature that memetic content on the internet is ephemeral and rapidly evolving~\cite{lakkaraju2013s, deza2015understanding, dubey2017modeling}, which makes the challenge of predicting the nature of content and the timeline of its existence an interesting problem. Building on this line of thought, we now describe the quantitative experiments we perform to assess our algorithm.
\subsection{Quantitative Evaluation}
\begin{table}
\centering
\small
\begin{tabular}{lcc}
\hline
\hline
\multirow{2}{*}{\textbf{Algorithm}} & \multicolumn{2}{c}{\textbf{Accuracy(\%)}} \\
& \textbf{Topic} & \textbf{Timeline} \\ \hline
SVM + Image Features \cite{deza2015understanding} & 41.09 & 24.14\\
SVM + 2x2 Dense HOG Features & 43.08 & 25.18 \\
Finetuned AlexNet \cite{krizhevsky2012imagenet} & 53.89 & 32.15 \\
Finetuned VGGNet16 \cite{simonyan2014very} & 57.21 & 33.02\\
Finetuned ResNet18\cite{simonyan2014very} & 58.17 & 35.16 \\  \hline
SVM + Word2VecPooling ~\cite{mikolov2013distributed} & 55.19 & 18.07\\
SVM + SkipThought Features ~\cite{kiros2015skip} & 58.81  & 19.15\\ \hline
SVM + ResNet18 + Word2VecPooling~\cite{mikolov2013distributed} & 65.35 & 32.15\\
SVM + ResNet18 + SkipThought Features~\cite{kiros2015skip} & 69.18 & 34.06\\ \hline
Xiao and Lee \cite{xiao2015discovering} & 70.57 & 38.18\\
Singh and Lee \cite{singh2016end} & 71.85 & 37.09\\
Dubey and Agarwal \cite{dubey2017modeling} & 75.57 & 40.17 \\ \hline
\textbf{Sparse Matching} (AlexNet, Word2VecPooling) & 72.18 & 37.15 \\
\textbf{Sparse Matching} (VGGNet16, Word2VecPooling) & 76.08 & 39.57 \\
\textbf{Sparse Matching} (ResNet18, Word2VecPooling) & 77.85 & 40.88 \\
\textbf{Sparse Matching} (AlexNet, SkipThought) & 74.23  & 43.44\\
\textbf{Sparse Matching} (VGGNet16, SkipThought) & 78.98  & 45.78 \\
\textbf{Sparse Matching} (ResNet18, SkipThought) & \textbf{80.69} & \textbf{46.91} \\ \hline \hline
\end{tabular}
\caption{On both tasks of topic and timeline prediction, we observe that Sparse Matching provides substantial improvements in performance.}
\label{table:content_prediction}
\end{table}
We perform a series of varied inference tasks to assess the quality of representations generated by our algorithm. The general framework for evaluation in these experiments is as follows - (i) we first select a CNN-based image feature extractor from the set of CNNs described earlier, and select one of two text feature extractors, (ii) the CNN feature extractor is fine-tuned naively on the target task using the assigned training set, and (iii) we extract both image features and text features from the designated neural networks, following the Sparse Matching algorithm. Finally, once the features are obtained, we train a multiclass SVM classifier~\cite{hsu2002comparison} for each of the following inference tasks. If the task is a ranking task (as Virality Prediction), we train a RankSVM~\cite{joachims2002optimizing} instead.\\
\begin{table*}
\centering
\small
\begin{tabular}{lccccc}
\hline
\hline
\multirow{2}{*}{\textbf{Algorithm}}& \multicolumn{5}{c}{{\textbf{Percentage Accuracy on Dataset}}} \\
& VCom\cite{deza2015understanding} & VPairs\cite{deza2015understanding} & VRPairs\cite{deza2015understanding}& MemeGenerator\cite{coscia2013competition} & QuickMeme\cite{coscia2014average} \\ \hline
RankSVM + Image Features \cite{deza2015understanding} & 53.40 & 61.60 & 58.49 & 59.12 & 57.05\\
RankSVM + 2x2 Dense HOG Features & 52.75 & 58.81 & 56.92 & 58.87 & 55.56 \\
RankSVM + AlexNet fc7 Features \cite{krizhevsky2012imagenet} & 54.41 & 61.93 & 58.58 & 59.91 & 58.03\\
RankSVM + VGGNet-16 fc7 Features \cite{simonyan2014very} & 55.18 & 63.01 & 59.15 & 60.12 & 61.12 \\ \hline
RankSVM + Word2VecPooling~\cite{mikolov2013distributed} & 60.11 & 61.78 & 59.94 & 61.49 & 62.02 \\
RankSVM + SkipThought~\cite{kiros2015skip} & 63.06 & 64.12 & 60.23 & 65.57 & 64.28 \\ \hline
RankSVM + ResNet18 + Word2VecPooling~\cite{mikolov2013distributed} & 66.05 & 70.98 & 70.33 & 71.06 & 69.45 \\
RankSVM + ResNet18 + SkipThought~\cite{kiros2015skip} & 69.36 & 74.51 & 72.09 & 75.53 & 73.81 \\ \hline
Xiao and Lee \cite{xiao2015discovering} & 63.25 & 75.23 & 73.22 & 70.11 & 71.21 \\
Singh and Lee \cite{singh2016end} & 65.87 & 76.20 & 74.38 & 72.25 & 70.08 \\
Dubey and Agarwal \cite{dubey2017modeling} & 68.09 & 78.38 & 76.95 & 74.43 & 74.51 \\ \hline
\textbf{Sparse Matching} (AlexNet, Word2VecPooling) & 70.02 & 82.21 & 80.04 & 79.53 & 79.95 \\
\textbf{Sparse Matching} (VGGNet16, Word2VecPooling) & 70.93 & 83.03 & 81.15 & 79.94 & 80.22 \\
\textbf{Sparse Matching} (ResNet18, Word2VecPooling) & 71.87 & 84.19 & 81.96 & 80.01 & 80.27 \\
\textbf{Sparse Matching} (AlexNet, SkipThought) & 70.86 & 83.05 & 81.29 & 80.25 & 80.16 \\
\textbf{Sparse Matching} (VGGNet, SkipThought) & 71.54 & 83.98 & 82.34 & 81.16 & 80.87 \\
\textbf{Sparse Matching} (ResNet18, SkipThought) & \textbf{73.03} & \textbf{85.15} & \textbf{82.62} & \textbf{81.80} & \textbf{80.9}1 \\
\hline \hline
\end{tabular}
\caption{\small Comparison of performance on the task of virality prediction. We observe that with our feature representation performance is unmatched compared to the existing state-of-the-art, by a significant margin.}
\label{table:virality_prediction}
\end{table*}
\textbf{Topic Prediction}: The Viral Images Dataset~\cite{deza2015understanding} is constructed of images scraped from Reddit~\cite{web_reddit}, along with the subreddits these images were posted to. The subreddits are curated lists belonging to a particular genre or topic of content, and example subreddits are \texttt{/r/funny, /r/aww, /r/wtf, /r/gaming} and \texttt{/r/atheism}, and the resulting inference task is that of predicting the correct topic or subreddit the image was posted to.\\ \hspace*{10pt}
The task is challenging primarily owing to the overlap in imagery across categories, and the overlap in context as well. For example, several categories may contain images of different animals or people, as well as a significant overlap in the template images that might be used. We compare our performance on this task with several benchmark and preivous state-of-the-art methods on this task, summarized in Table~\ref{table:content_prediction}. Since there are 20 different categories, random chance performs at 5\%. Due to the immense overlap in content, naive algorithms perform poorly, with less than 50\% prediction accuracy.\\ \hspace*{10pt}
By just considering the text content itself, we see a slight increase in performance, which increases further still when both feature modalities are combined (image CNN is fine-tuned). However, as hypothesized, these features are corrupted by the coupling of text and image content, and hence perform averagely. Approaches that take context into account, such as the work of Singh and Lee~\cite{singh2016end}, and Dubey and Agarwal~\cite{dubey2017modeling} perform much better, however, they face challenges in decoupling text information. Finally, we see that Sparse Matching performs substantially better, with a maximum gain (with ResNet18 visual features and Skip Thought language features), of $5.12\%$ in performance.\\ \hspace*{10pt}
\textbf{Timeline Prediction}: The results of the preivous experiment provide us with evidence regarding the representational power of our method. An extension of the first experiment would be to predict, along with the topic of the content, the time period it was posted in. Applications of such a task are present in content creation, and prediction of popularity trends in social media. Since content on viral image websites and memes are continually evolving, we see that the content alone can be a strong indicator of the time period it represents, as displayed by our prediction results in Column 2 of Table~\ref{table:content_prediction}.\\
We observe that naive baselines for both basic image features and text features fail miserably at this task, capturing very little information about the context of an image, crucial for a task like timeline prediction. Deep image features perform slightly better, and a combination of deep image features and text features provides good performance, which is ousted by previous virality prediction techniques, and Sparse Matching. The decoupled context provides a signficant boost over the previous state-of-the-art, with an improvement of $\textbf{6.74\%}$.\\ \hspace*{10pt}
\textbf{Virality Prediction}: With the impressive increase in performance on timeline prediction, we have evidence supporting the improved ability of our method to identify temporal distinctions in memetic imagery. This demonstration leads us to our final quantitative experiment, virality prediction. As summarized by recent work on detection of viral content and cascades in social networks~\cite{lakkaraju2013s}, capturing virality is a difficult problem, with challenges arising from the presence short-lived content, rapidly evolving diverse imagery. With our model, however, we hypothesize that predicting virality would be possible with greater efficiency owing to the stronger, robust representation.\\ \hspace*{10pt}
Hence, we evaluate the prediction performance on the primary inference task of virality prediction, following the methodology introduced in~\cite{dubey2017modeling}. Here, the prediction of virality is taken to be a pairwise classification task, where given an input pair of images, the algorithm needs to predict the more viral image of the two. This evaluation scheme has been shown to be robust to network effects~\cite{dubey2017modeling}, and is representative of evaluation of the image virality based on the content alone. We follow the same procedure, replacing our SVM classifier with a Ranking SVM~\cite{joachims2002optimizing} classifier to learn a function that provides pairwise predictions. We evaluate our algorithm on 5 different datasets  - the 3 test splits of the Viral Images Dataset~\cite{deza2015understanding} - (i) Viral-Complete, (ii) Viral Pairs, and (iii) Viral-RandomPairs, and the Memegenerator~\cite{coscia2013competition} and Quickmeme datasets~\cite{coscia2014average}. For all experiments where a train-test split is not predefined, we use a 70-20-10 (train-test-val) split, following standard protocol.\\
Our experiment is summarized in Table~\ref{table:virality_prediction}. Note here that since this is a pairwise prediction task, random chance performs at 50\%. Again, as seen in the previous experiments, naive image-based algorithms perform only slightly better than random chance, with accuracies around 50-60\%. Operating only on extracted text data gives us slightly better performance, but it fails to generalize since a lot of imagery do not possess any relevant text at all, which provides the algorithms with only noise. Sparse Matching provides large improvements, as high as \textbf{7.37\%} in some cases. Even the weakest versions of Sparse Matching perform \textbf{3.8\%} better than the previous state-of-the-art.\\ \hspace*{10pt}
Finally, we attempt to analyse the durability of our method in virality prediction. Since the nature of viral content changes rapidly over time, we would want to learn a classifier that can perform well for longer periods of time, given an input training set taken at any prior time. We  would not want to keep re-training our algorithm often as new data keeps coming in, and reduce the time spent updating our model. To ascertain the invariance of methods in such a manner, we devise a final quantitative experiment.\\
\textbf{Temporal Robustness Analysis}: To examine the resilience of our algorithm to evolving data, we devise the following experiment: The Memegenerator dataset~\cite{coscia2013competition} contains images from 13 different time periods, as described earlier. We select data from the $i^{th}$ time period as our training data, and test the trained algorithm on the $k-i$ sets of test images from subsequent time periods. We perform $\sum_{i=1}^{k-1} (k-i) = \frac{k(k-1)}{2}$ total experiments, which in our case is 78. We report the average obtained performance for each algorithm, summarized in Table~\ref{table:temporal_analysis}.\\
\begin{table}
\centering
\small
\begin{tabular}{lc}
\hline
\hline
\textbf{Algorithm} & \textbf{Accuracy(\%)} \\ \hline
SVM + Image Features \cite{deza2015understanding} & 51.02\\
SVM + 2x2 Dense HOG Features & 51.18  \\
Finetuned AlexNet \cite{krizhevsky2012imagenet} & 51.75 \\
Finetuned VGGNet16 \cite{simonyan2014very} & 51.56 \\
Finetuned ResNet18 \cite{simonyan2014very} & 51.97 \\  \hline
SVM + Word2VecPooling ~\cite{mikolov2013distributed} & 52.02 \\
SVM + SkipThought Features ~\cite{kiros2015skip} & 52.95 \\ \hline
SVM + Finetuned ResNet18 + Word2VecPooling~\cite{mikolov2013distributed} & 55.68 \\
SVM + Finetuned ResNet18 + SkipThought Features~\cite{kiros2015skip} & 56.35 \\ \hline
Xiao and Lee \cite{xiao2015discovering} & 62.67 \\
Singh and Lee \cite{singh2016end} & 61.98 \\
Dubey and Agarwal \cite{dubey2017modeling} & 62.83 \\ \hline
\textbf{Sparse Matching} (AlexNet, Word2VecPooling) & 61.78  \\
\textbf{Sparse Matching} (VGGNet16, Word2VecPooling) & 61.97  \\
\textbf{Sparse Matching} (ResNet18, Word2VecPooling) & 63.49  \\
\textbf{Sparse Matching} (AlexNet, SkipThought) & 64.07  \\
\textbf{Sparse Matching} (VGGNet16, SkipThought) & 64.18  \\
\textbf{Sparse Matching} (ResNet18, SkipThought) & \textbf{65.12} \\
\hline \hline
\end{tabular}
\caption{Evaluation of Sparse Matching on the Temporal Robustness task.}
\label{table:temporal_analysis}
\end{table}
It is vital to understand the difficulty of this task compared to the previous virality prediction experiment. For each time period, we have (on average) only 7\% of the previously available training data, which provides a very sparse signal. This is consistent with the observation that most methods (including naive deep learning algorithms) perform very close to random chance, given the lower correlation between the observed training data and the test data. We still observe, however, that specialized methods perform much better than random chance, and Sparse Matching provides the best performance, showcasing again the richness and versatility of the method in capturing context.
\section{Discussion and Future Work}
\begin{figure}[b]
  \centering
  \includegraphics[width=0.45\textwidth]{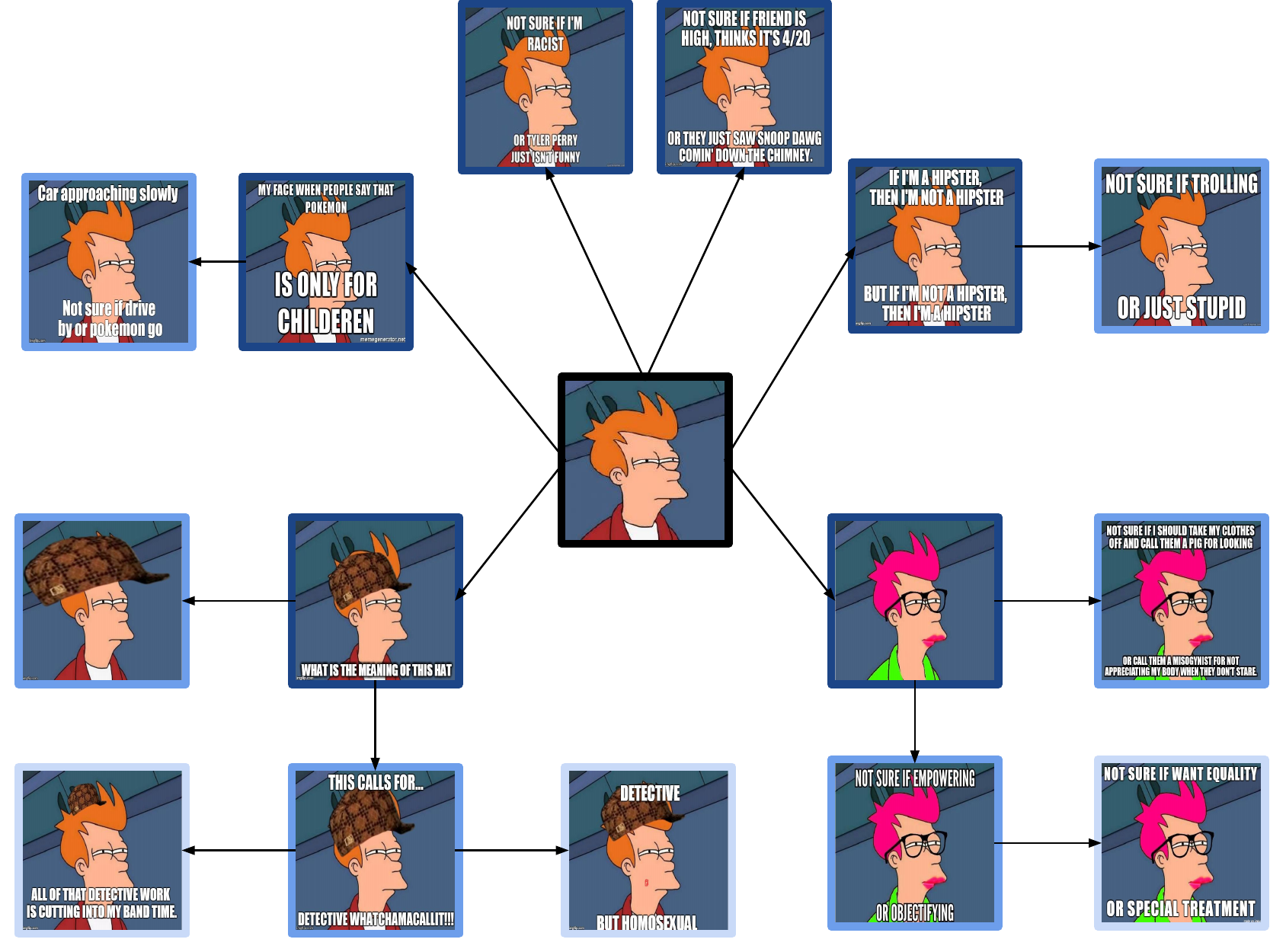}
  \caption{Pruned phylogenetic tree constructed from 1 meme template and its neighbours using distances generated by Sparse Matching.}
	\label{fig:phylogenetic_tree}
\end{figure}

In this work, we introduced the task of understanding the creation, mutation and propagation of image macros online as a representation for the growth and evolution of a meme. We introduced a method that succesfully maps any arbitrary image macro to the template it was created from, and then embeds it into a powerful semantic space. We demonstrated the robustness and richness of the semantic embedding space via a variety of relevant experiments on online imagery - image retrieval, clustering, content analysis, popularity analysis and temporal evolution, all with remarkable performance.\\ \hspace*{10pt}
Our work is the first of its kind in the domain of web content analysis that looks at the virality prediction problem through the lens of the Image Macro structure, and successfully exploits it. While prior work in the domain of virality prediction from content has focused on the originality of memes~\cite{coscia2014average}, popularity-promoting regions within meme images~\cite{dubey2017modeling}, relative attributes and category analysis~\cite{deza2015understanding}, our approach provides a technique to systematically study the evolution of ideas and memes online.\\\hspace*{10pt}
As a demonstration of this technique, we provide a sample phylogenetic tree constructed from the pairwise distances between different image macros, obtained using our algorithm (see Figure~\ref{fig:phylogenetic_tree}). This tree displays the mutation of memes from a popular template, created using MATLAB's \texttt{seqlinkage} command (the figure displays selected nodes). We see that our embedding is powerful enough to capture subtle evolutionary patterns in the mutation of the meme, with children being semantically linked either through imagery (note the subtree with the hat overlay), or through text (subtree with pokemon text). With large amounts of time-series meme imagery, our embedding technique paves way for the study of systematic evolution of ideas on the Internet.\\ \hspace*{10pt}
All previous efforts in the study of image virality have been futile at extracting these subtle variations in memes, and creating evolutionary trees such as the one displayed, despite having success at the task of predicting virality~\cite{dubey2017modeling, deza2015understanding}. Our contribution through this work enables the large-scale study and analysis of such evolutionary trends online, with numerous applications in combining the effects of content and network structure in understanding information diffusion and evolution. Such a study can be influential in creating methods for identifying original content and their sources, and creating a robust science for understanding how multimedia propagates on the Internet.
\bibliographystyle{ACM-Reference-Format}
\bibliography{bib_new}

\end{document}